\documentclass[twocolumn]{article}

\usepackage{PRIMEarxiv}

\usepackage[utf8]{inputenc} 
\usepackage[T1]{fontenc}    
\usepackage{url}            
\usepackage{booktabs}       
\usepackage{amsfonts}       
\usepackage{nicefrac}       
\usepackage{microtype}      
\usepackage{lipsum}
\usepackage{fancyhdr}       
\usepackage{graphicx}       

\usepackage{cite}
\usepackage{amsmath,amssymb,amsfonts}
\usepackage{algorithmic}
\usepackage{graphicx}
\usepackage{algorithm,algorithmic}
\usepackage[hidelinks]{hyperref}
\usepackage[capitalise]{cleveref}

\usepackage{caption}
\usepackage{subcaption}

\usepackage{booktabs}
\usepackage{multirow}
\usepackage{amsmath}

\usepackage{enumitem}

\graphicspath{{media/}}     

\title{Advancing Cell Detection in Anterior Segment Optical Coherence Tomography Images}

\author{
  Boyu Chen\\
  Institute of Health Informatics \\
  University College London \\
  \texttt{boyu.chen.19@ucl.ac.uk} \\
   \And
  Ameenat L. Solebo\\
  Great Ormond Street Institute of Child Health \\
  University College London \\
  \texttt{a.solebo@ucl.ac.uk} \\
  \And
  Paul Taylor\\
  Institute of Health Informatics \\
  University College London \\
  \texttt{p.taylor@ucl.ac.uk} \\
}

\begin{document}
\twocolumn[
\maketitle
\begin{abstract}
Anterior uveitis, a common form of eye inflammation, can lead to permanent vision loss if not promptly diagnosed. Monitoring this condition involves quantifying inflammatory cells in the anterior chamber (AC) of the eye, which can be captured using Anterior Segment Optical Coherence Tomography (AS-OCT). However, manually identifying cells in AS-OCT images is time-consuming and subjective. Moreover, existing automated approaches may have limitations in both the effectiveness of detecting cells and the reliability of their detection results. To address these challenges, we propose an automated framework to detect cells in the AS-OCT images. This framework consists of a zero-shot chamber segmentation module and a cell detection module. The first module segments the AC area in the image without requiring human-annotated training data. Subsequently, the second module identifies individual cells within the segmented AC region. Through experiments, our framework demonstrates superior performance compared to current state-of-the-art methods for both AC segmentation and cell detection tasks. Notably, we find that previous cell detection approaches could suffer from low recall, potentially overlooking a significant number of cells. In contrast, our framework offers an improved solution, which could benefit the diagnosis and study of anterior uveitis. Our code for cell detection is publicly available at: \href{https://github.com/joeybyc/cell_detection}{https://github.com/joeybyc/cell\_detection}.

\keywords{Anterior Segment Optical Coherence Tomography \and Anterior Uveitis \and Medical Image Segmentation \and Medical Image Object Detection}
\end{abstract}
\vspace{1cm}
]

\section{Introduction}
\label{sec:introduction}
Anterior uveitis is a common inflammatory eye condition that causes significant visual impairment worldwide, especially among children \cite{wakefield2005epidemiology, angeles2015characteristics, maleki2022pediatric}. If not promptly diagnosed and treated, this condition can lead to ocular tissue remodeling and potentially permanent vision loss \cite{agrawal2012cataract, lee2014autoimmune}. Thus, early detection and intervention are crucial for managing and preventing most uveitis-related complications \cite{keane2014objective}.

The diagnostic hallmark of anterior uveitis is the presence of inflammatory cells in the eye's anterior chamber (AC) \cite{standardization2005standardization}. Traditionally, clinicians rely on slit-lamp biomicroscopic ocular examination to identify and count these cells. They then convert the count to the internationally accepted Standardization of Uveitis Nomenclature (SUN) clinical grading scale \cite{standardization2005standardization}, which indicates the severity of inflammation. Nevertheless, this approach has several limitations, including subjectivity, variability between examiners, and insensitivity to changes in inflammation severity \cite{kempen2008interobserver, maring2022grading, wong2009effect}. Although quantitative assessment of 'flare' (proteinaceous change in the aqueous humor, another clinical feature in some forms of uveitis) is possible using a laser flare photometer, there are currently no commercially available devices to quantify AC cells \cite{keino2022automated}.

To tackle this gap, Anterior Segment Optical Coherence Tomography (AS-OCT) has recently emerged as a promising diagnostic and monitoring tool for anterior uveitis \cite{keino2022automated, igbre2014high, kumar2015analysis, invernizzi2017objective, akbarali2021imaging, tsui2022quantification, solebo2023anterior, lu2020quantitative, etherton2023quantitative, sharma2015automated, agarwal2009high, rose2015aqueous, baghdasaryan2019analysis, kang2021development, sorkhabi2022assessment}. This technique captures high-resolution cross-sectional images of the front of the eye, where cells appear as hyper-reflective white particles. Identifying and counting these particles can provide a quantitative assessment of the severity of inflammation. However, manual analysis (cell identification) of AS-OCT by clinicians is time-consuming and subjective.

Although automated identification is beneficial, previous approaches \cite{keino2022automated, lu2020quantitative, etherton2023quantitative, sharma2015automated, agarwal2009high, rose2015aqueous, baghdasaryan2019analysis, kang2021development, sorkhabi2022assessment} are open to limitations in effectiveness and reliability. Most of them rely on thresholding algorithms (detailed in \cref{subsubsec:thresholding_algorithm}), which convert an image into a binary mask and count the number of connected components in the mask as the number of cells. In our study, we demonstrate that the commonly used thresholding algorithm for cell detection suffers from low recall, potentially overlooking a significant number of cells and leading to suboptimal effectiveness.

Moreover, previous studies evaluated their detection results by modeling the automated cell count against the slit lamp SUN score or a human expert's manual AS-OCT derived cell count, rather than assessing the correctness of each detected cell's location. In other words, they examined the correlation between the automated cell count and these reference values, but did not verify whether each detected object corresponds to an actual cell in the image. This could negatively affect the reliability.

To address these problems, we propose an automated framework called the Anterior Chamber Cell Detector (ACCDor), comprising a zero-shot chamber segmentation module (CSM) and a cell detection module (CDM). The CSM segments the AC area without relying on training data and the CDM identifies the cells within the segmented area. Unlike previous studies, we treat cell identification as an object detection task and evaluate the detection using cell location annotations. Our experiments demonstrate that ACCDor can outperform current state-of-the-art (SOTA) methods for both AC segmentation and cell detection tasks.

Our study makes the following contributions:

1) We propose a fully automated framework (ACCDor) for detecting inflammatory cells from AS-OCT images of anterior uveitis patients, comprising CSM and CDM.

2) CSM segments the AC area in a zero-shot manner, providing a more efficient and practical solution for clinical applications. It outperforms SOTA segmentation models trained on annotated datasets, achieving an Intersection over Union of 96.41\% and a Dice Coefficient of 98.21\%.

3) CDM localizes individual cells within the segmented AC area, achieving a precision, recall, and F1-score of 86.88\%, 87.23\%, and 87.02\%, respectively. In comparison, the baseline with the highest F1-score achieves 93.59\%, 70.58\%, and 80.47\% for these metrics. The F1-score, which balances precision and recall, is 6.55\% higher for CDM, demonstrating its superior overall performance. Although the precision of CDM is 6.71\% lower than the baseline, the recall increases substantially by 16.65\%, indicating that CDM can identify a significantly higher number of cells.

4) Our experiments demonstrate that cell identification through the commonly-used thresholding algorithm can lead to a relatively low recall, potentially overlooking a significant number of cells in the images. Although previous studies based on thresholding have reported a strong correlation between identified cell counts and the SUN score, this correlation may have been constrained by incomplete cell detection. This finding could provide a valuable perspective for future studies.

The paper is organized as follows: \cref{sec:related_works} describes how previous studies identify cells from AS-OCT images. \cref{sec:method} presents ACCDor, our proposed framework. \cref{sec:experiment} details the experiments, while \cref{sec:discussion} discusses the results. Finally, \cref{sec:conclusion} summarizes the paper.

\section{Related Works}
\label{sec:related_works}
Identifying and counting inflammatory cells within AS-OCT images is important for the diagnosis and study of anterior uveitis. We categorize approaches to this problem as manual, semi-automated, and fully automated.

Manual identification, where a clinician visually inspects the image and marks cells \cite{igbre2014high, kumar2015analysis, invernizzi2017objective, akbarali2021imaging, tsui2022quantification, solebo2023anterior}, is straightforward but time-consuming and subjective. To reduce manual effort, semi-automated techniques have been explored \cite{lu2020quantitative, etherton2023quantitative}. In these studies, clinicians first manually delineate the AC region, and then apply the default thresholding algorithm in ImageJ\footnote{https://imagej.net/ij/}, a commonly-used tool for image processing, to detect cells within that region. While this reduces burden compared to fully manual approaches, outlining the AC still requires extra time from clinicians.

Fully automated approaches aim to eliminate manual effort and can be further divided into thresholding-based and deep learning-based methods. Thresholding-based methods \cite{keino2022automated, sharma2015automated, agarwal2009high, rose2015aqueous, baghdasaryan2019analysis, kang2021development} identified cells by applying thresholding algorithms (detailed in \cref{subsubsec:thresholding_algorithm}) directly to the image. However, most of them did not precisely identify the AC area and instead cropped out a rectangular region partially containing the AC. The cropped region may include unrelated elements from outside the AC or miss relevant cells within the AC itself. This can negatively affect the cell detection.

Moreover, these methods often evaluate detection results by modeling the detected cell counts against slit lamp SUN scores or manual AS-OCT cell counts, without verifying the correctness of each detected object's location, potentially impacting the reliability of their results. Furthermore, our study finds that the widely-used thresholding algorithm for cell detection tends to have relatively low recall, potentially overlooking a significant number of cells.

Deep learning offers an alternative approach to solve the problem. Sorkhabi et al. \cite{sorkhabi2022assessment} trained a UNet \cite{ronneberger2015u} to segment the AC and then trained another UNet to detect the cells, with the training data consisting of cropped regions from AS-OCT images. While this study reported a strong correlation between the number of detected cells and the SUN score, the precision and recall of cell detection were suboptimal.

To address the limitations of previous approaches, ACCDor differs in several aspects. First, CSM is designed to automatically and accurately delineate the AC area. It can achieve this without relying on training data, in contrast to earlier studies that either depend on manual drawing or employ deep learning models trained on annotated datasets. Second, CDM treats cell identification as an object detection task, directly localizing individual cells within the segmented AC area. This approach offers a more reliable and accurate assessment of cell detection performance compared to previous studies. By focusing on these innovations, ACCDor can provide a more reliable and efficient solution for cell identification in AS-OCT images.

\section{Method}
\label{sec:method}
We propose ACCDor for identifying cells in AS-OCT images, which consists of a zero-shot chamber segmentation module (CSM) and a cell detection module (CDM). CSM delineates the AC area by leveraging the power of a foundation model, Segment Anything Model (SAM) \cite{kirillov2023segment}. This module can automatically generate prompts and feed them into the pre-trained SAM model along with the input image to obtain the AC mask. This focuses cell detection within the relevant area, eliminating false positives outside the AC. Subsequently, CDM localizes the cells within this area by modifying a thresholding algorithm to detect more objects and employing a cell classifier to filter out false positive detections.

\subsection{Preliminaries}
\subsubsection{Segment Anything Model (SAM)}
SAM \cite{kirillov2023segment} is a SOTA foundation model that has shown impressive performance in segmenting images. To use SAM, users first provide prompts, such as points clicked within the object of interest (point prompts) or bounding boxes drawn around the object (box prompts). These prompts are then fed into the pre-trained SAM model along with the input image. SAM then generate a mask that accurately delineates the object of interest, ensuring that the mask includes the point prompts or falls within the box prompts.

\subsubsection{Thresholding Algorithm}
\label{subsubsec:thresholding_algorithm}
Thresholding algorithms are simple yet effective techniques for image segmentation. To use them to detect cells, an image is first converted to grayscale, where pixel intensities range from 0 (pure black) to 255 (pure white). Then, a threshold is used to split the image into a binary mask with two regions: pixels with values above the threshold are considered to represent cells, while pixels below the threshold are treated as background. The threshold can be either a pre-defined constant or automatically calculated from thresholding algorithms. Within the mask, the number of connected components is counted as the number of cells.

The Isodata algorithm \cite{ridler1978picture} is the default thresholding method in ImageJ and is commonly used for cell detection in AS-OCT images \cite{lu2020quantitative, etherton2023quantitative, agarwal2009high, baghdasaryan2019analysis}. It begins with an initial threshold (e.g., the mean intensity of the image), and iteratively calculates the average pixel intensities of the two regions (lower and upper class means). The new threshold is then set to the average of these class means, and the process is repeated until the threshold converges, resulting in the optimal threshold to separate an image into background and object regions.

\begin{figure}[b]
\centerline{\includegraphics[width=\columnwidth]{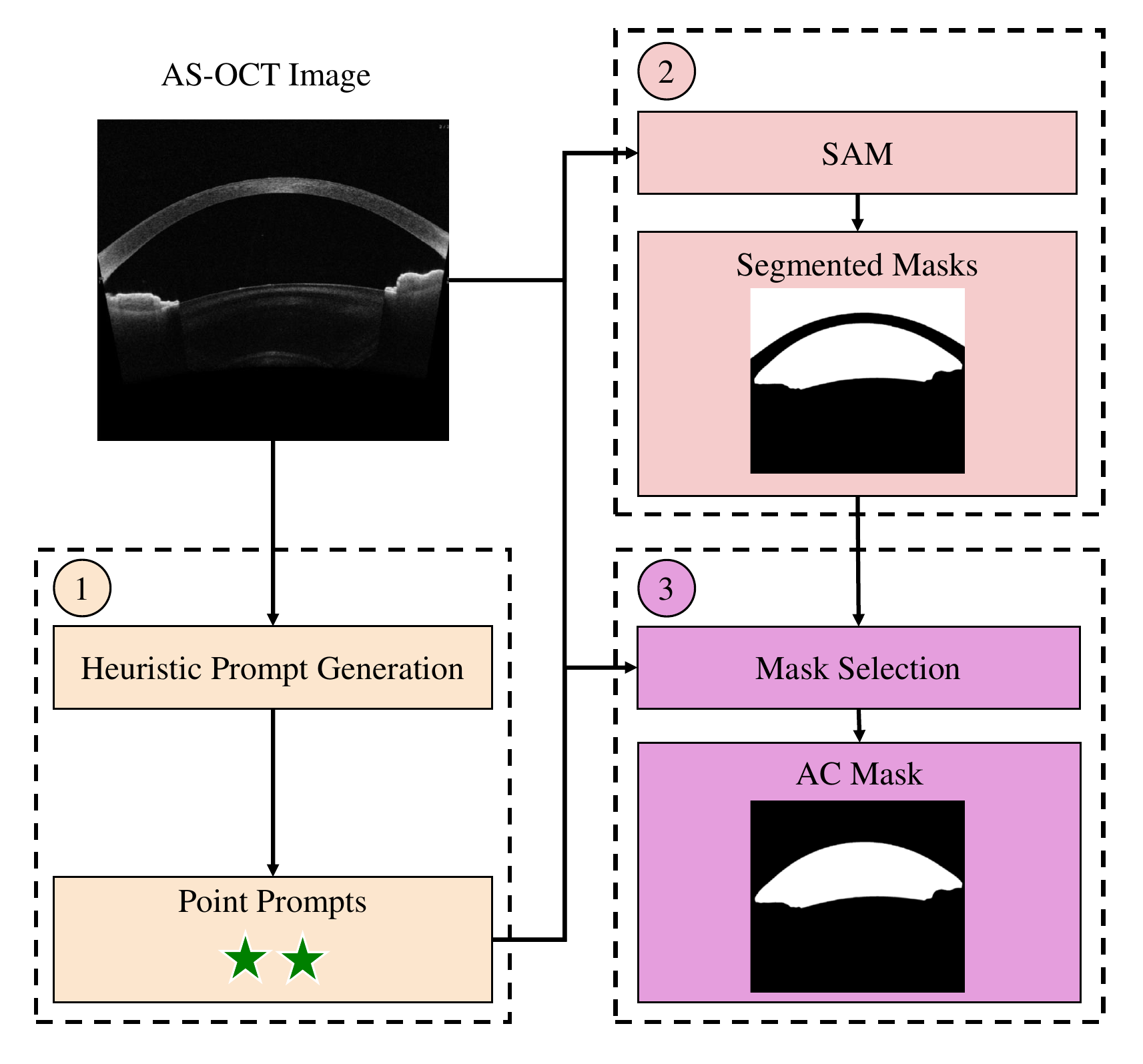}}
\caption{The architecture of CSM.}
\label{fig:CSM_architecture}
\end{figure}

\begin{figure}
\centering
\begin{subfigure}[b]{0.15\textwidth}
    \includegraphics[width=\textwidth]{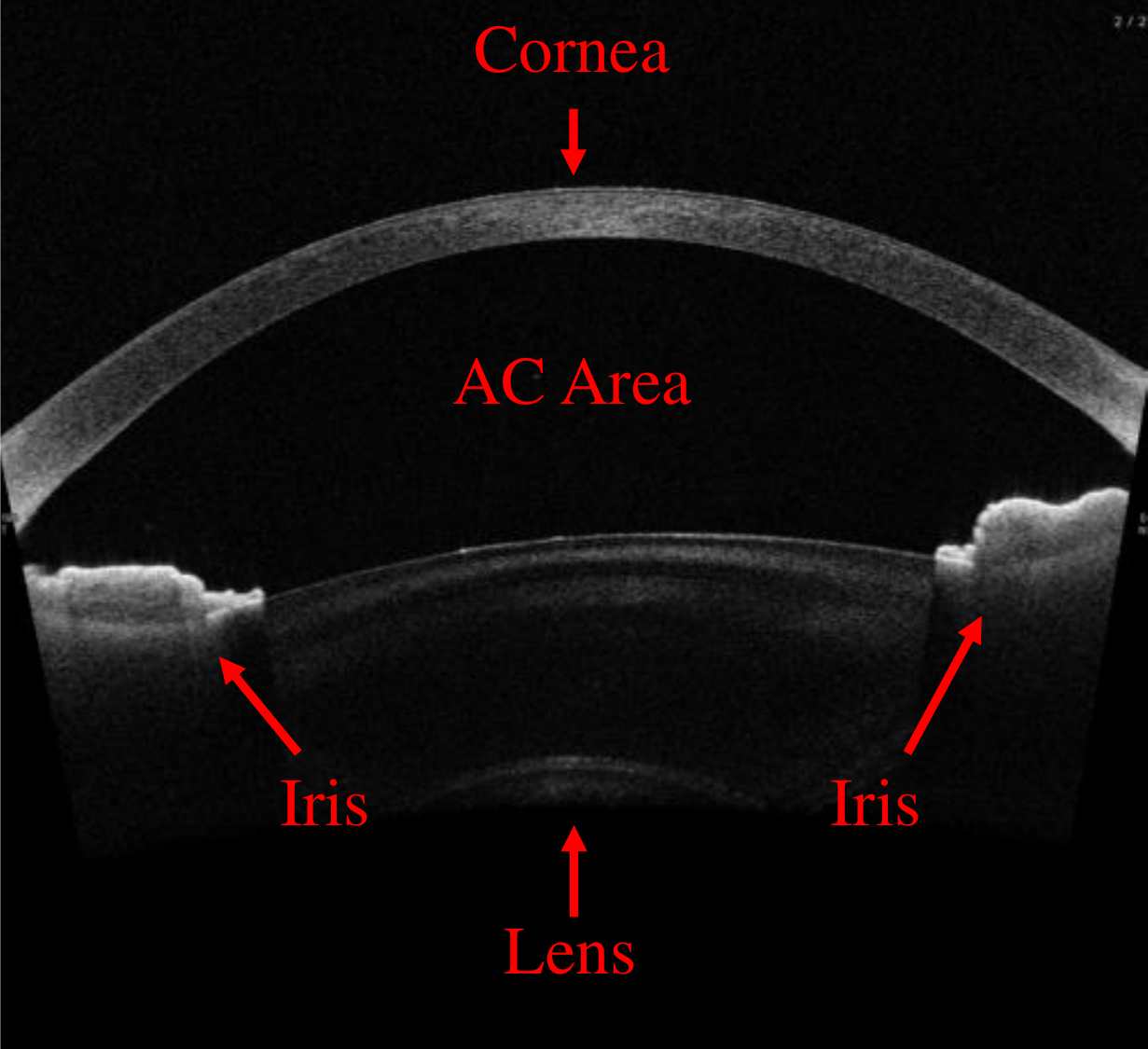}
    \caption{}
    \label{fig:CSM_a}
\end{subfigure}
\hfill
\begin{subfigure}[b]{0.15\textwidth}
    \includegraphics[width=\textwidth]{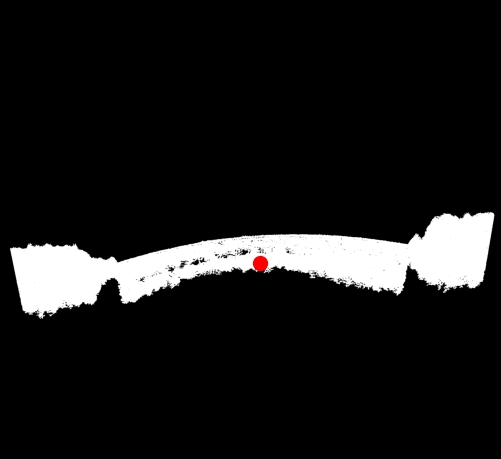}
    \caption{}
    \label{fig:CSM_b}
\end{subfigure}
\hfill
\begin{subfigure}[b]{0.15\textwidth}
    \includegraphics[width=\textwidth]{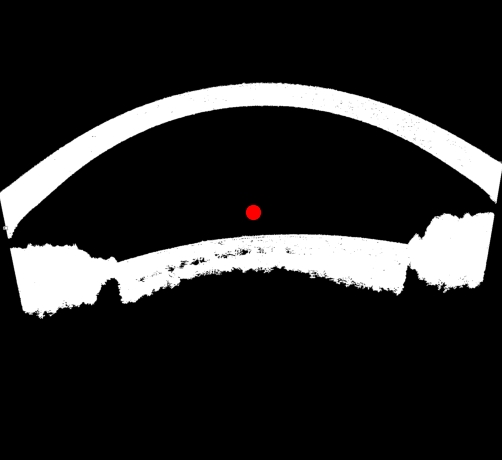}
    \caption{}
    \label{fig:CSM_c}
\end{subfigure}
\vspace{1em}

\begin{subfigure}[b]{0.15\textwidth}
    \includegraphics[width=\textwidth]{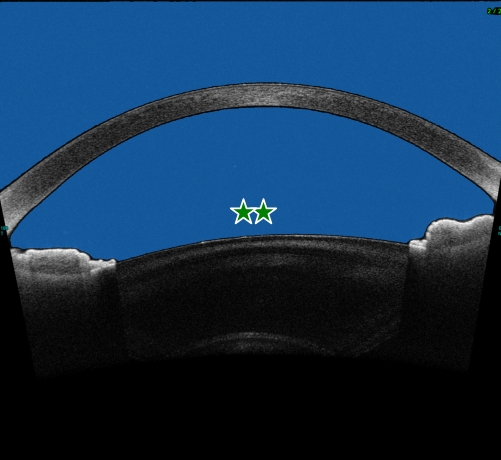}
    \caption{}
    \label{fig:CSM_d}
\end{subfigure}
\hfill
\begin{subfigure}[b]{0.15\textwidth}
    \includegraphics[width=\textwidth]{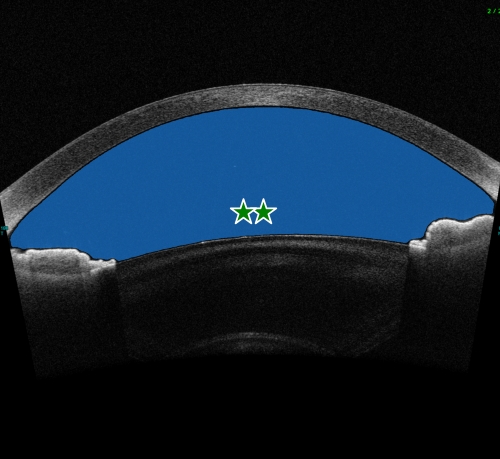}
    \caption{}
    \label{fig:CSM_e}
\end{subfigure}
\hfill
\begin{subfigure}[b]{0.15\textwidth}
    \includegraphics[width=\textwidth]{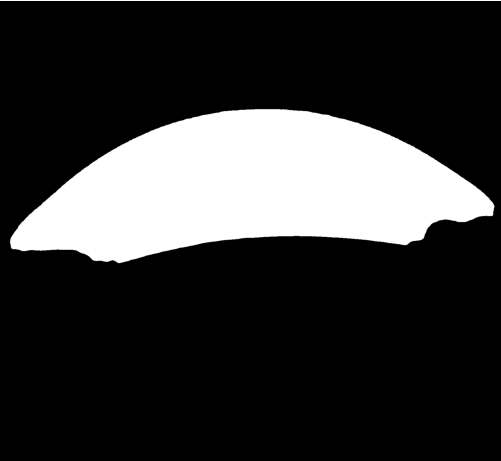}
    \caption{}
    \label{fig:CSM_f}
\end{subfigure}

\caption{Illustration of the CSM for AC segmentation.}
\label{fig:CSM}
\end{figure}

\subsection{Zero-Shot Chamber Segmentation Module (CSM)}
We treat the delineation of the AC area in AS-OCT images as an image segmentation task. To achieve this, CSM employs a three-step strategy, as illustrated in \cref{fig:CSM_architecture}:
\begin{enumerate}[leftmargin=1em, itemsep=0em]
\item  Generate point prompts for the input AS-OCT image.
\item Feed the prompts and the image into the pre-trained SAM model to obtain segmented masks.
\item Select the final AC mask from the segmented masks.
\end{enumerate}
To segment the AC area using SAM, the generated point prompts should fall within the AC region. We hypothesize that the centroid of the anterior segment lies within the AC area, since the anterior segment (which comprises the cornea, iris, and lens) surrounds the AC area (\cref{fig:CSM_a}). We design a Heuristic Prompt Generation (HPG) algorithm, which automatically generates point prompts near this centroid. The HPG follows five steps:
\begin{enumerate}[leftmargin=1em, itemsep=0em]
\item Convert the input AS-OCT image to grayscale.
\item Calculate the mean pixel intensity of the image and use it as a cutoff to create a binary mask: pixels with intensity lower than the cutoff are considered background, while those with higher intensity are treated as objects.
\item Treat the largest object in the binary mask as the anterior segment mask. In some cases, the anterior segment appears split into two (\cref{fig:CSM_c}), leading to an incomplete mask and an erroneous centroid (red point in \cref{fig:CSM_b}). To address this, HPG checks if the ratio of the area of the second-largest object to the largest object's area exceeds a threshold $R_{AS}$. If this condition is met, HPG considers the second-largest piece as part of the anterior segment mask (\cref{fig:CSM_c}).
\item Calculate the centroid ($X_{centroid}$, $Y_{centroid}$) of the mask (red point in \cref{fig:CSM_c}). In the image coordinate system of this study, the top-left corner is the origin (0, 0), with $X_{centroid}$ increasing when moving from top to bottom and $Y_{centroid}$ increasing when moving from left to right. The centroid is calculated based on:
\begin{equation} \nonumber
(X_{centroid}, Y_{centroid}) = (\frac{\sum_{i=1}^{N_{AS}} x_i}{N_{AS}}, \frac{\sum_{i=1}^{N_{AS}} y_i}{N_{AS}})
\end{equation}
where ($x_{i}$, $y_{i}$) denotes the coordinates of the point within the anterior segment mask and $N_{AS}$ is the total number of points within the mask.
\item Generate $N_p$ point prompts, $p_i$ ($i \in [1, N_p]$), using the formula $p_i = (X_{centroid} + \lambda^X_i, Y_{centroid} + \lambda^Y_i)$.
\end{enumerate}
After generating the prompts, CSM feeds them and the input image into the pre-trained SAM model to obtain the segmented areas. As illustrated in \cref{fig:CSM_d,fig:CSM_e}, the green stars indicate the generated prompts, while the blue regions represent the areas segmented by SAM. In certain cases, SAM may segment additional areas that are not part of the AC, such as the upper region in \cref{fig:CSM_d}. To address this, CSM selects only the area containing the prompts as the final AC mask, as shown in \cref{fig:CSM_f}. This mask is then used in the CDM to ensure that only detected cells within the AC are considered for further analysis.

\begin{figure*}
\centerline{\includegraphics[width=\textwidth]{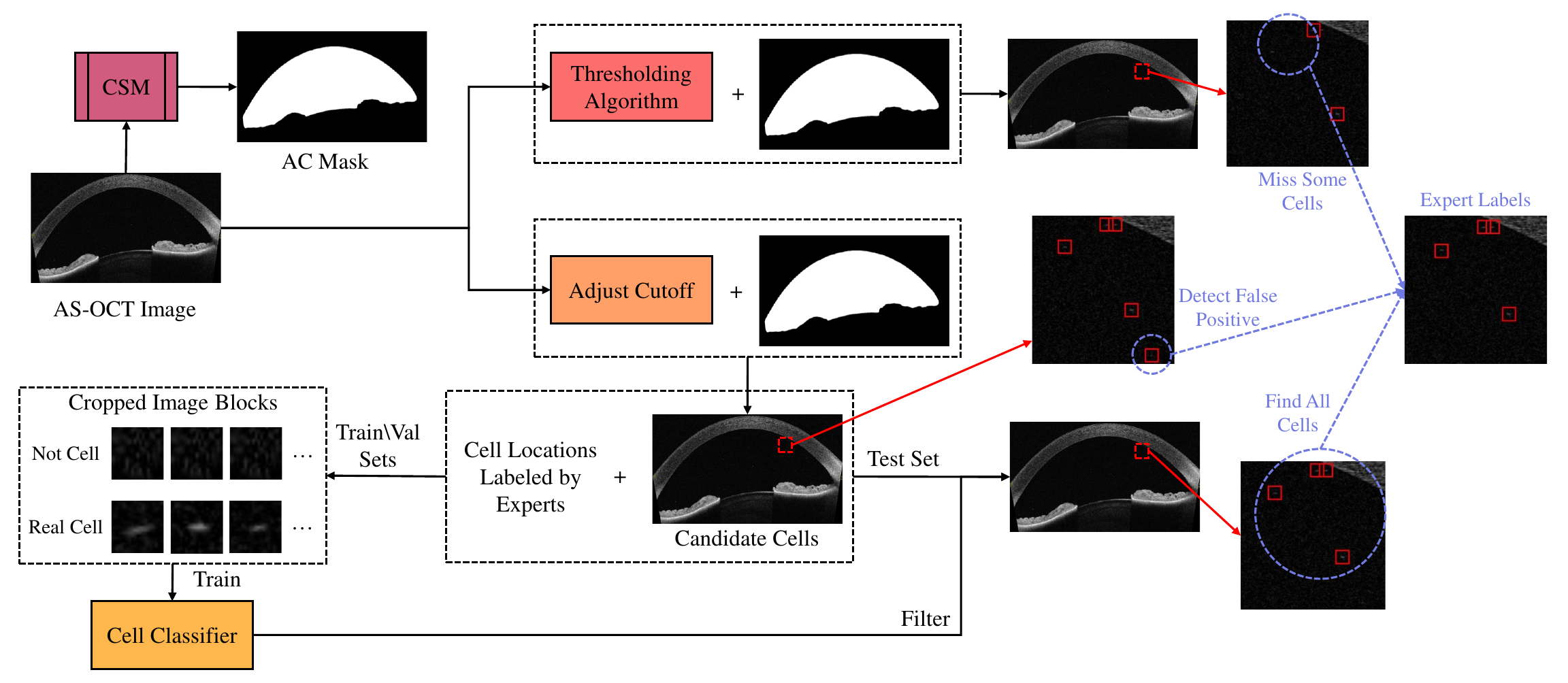}}
\caption{The architecture of CDM.}
\label{fig:CDM_architecture}
\end{figure*}

\begin{figure}
\centerline{\includegraphics[width=\columnwidth]{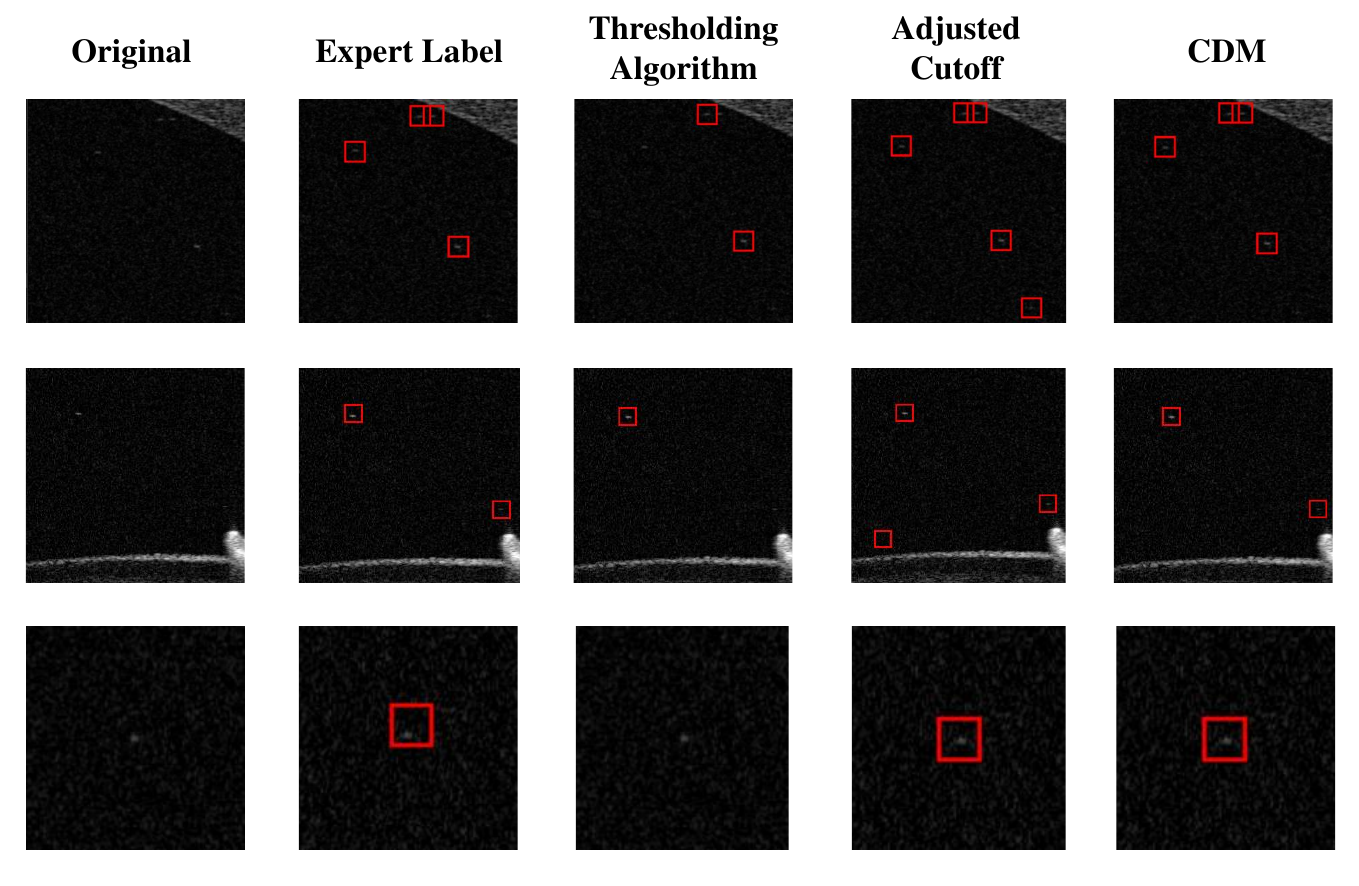}}
\caption{Visualization of the cell identification results.}
\label{fig:CDM_process}
\end{figure}

\subsection{Cell Detection Module (CDM)}
We treat cell identification as an object detection task, aiming to identify each cell with a $h \times w$ bounding box in the AS-OCT image with dimensions of $H \times W$ pixels. This task is challenging as the size of a cell is extremely small. To address this, we develop CDM (Fig.\ref{fig:CDM_architecture}). We first modify a thresholding algorithm to detect more objects (candidate cells) and then use a cell classifier to filter candidate cells by discarding false positive detections.

We design an adjusted cutoff algorithm, which modifies a base thresholding algorithm. The motivation is that thresholding algorithms may miss a number of cells, as the detection is sensitive to the threshold value. A higher threshold may cause fewer cells to be considered as objects since cells with pixel intensities below the threshold will be classified as background. Our proposed algorithm adjusts the threshold computed from a base thresholding algorithm, multiplying it by a factor $\alpha$ ($\alpha \in (0, 1)$). A lower $\alpha$ results in a lower threshold, allowing more objects with lower pixel intensities to be included, thereby improving the detection of cells that might have been missed by the original threshold.

Based on the new threshold, the image is converted into a binary mask containing objects and background. Previous expert annotations indicate that objects that are very small or large are unlikely to be cells \cite{lu2020quantitative, etherton2023quantitative}, so after thresholding, objects with an area outside the range [$\beta_{min}$, $\beta_{max}$] are discarded. Additionally, objects located outside the AC mask from CSM are removed. The centroid of each remaining isolated object is then treated as the center of a predicted bounding box, representing a detected object.

While the adjusted cutoff may identify more cells, it may also include more false positives (e.g., noise). To mitigate this, we treat every detected bounding box as a candidate cell and train a neural network (cell classifier) to distinguish "real cell" from "not cell". We split the images into training, validation, and testing sets. For the training set, we apply the adjusted cutoff to obtain the objects and crop a $h \times w$ pixel area from the original image based on each detected bounding box. The cropped image is labeled as "real cell" if it contains a coordinate label indicating the presence of a cell according to human expert consensuss; otherwise, it is labeled as "not cell".

To determine the optimal $\alpha$ value for the adjusted cutoff, CDM searches a range from $\tau_{min}$ to $\tau_{max}$ with step size $\tau_{step}$. For each $\alpha$, CDM applies the adjusted cutoff to obtain predicted boxes from both training and validation sets. It then trains a cell classifier using cropped images from the training set's predicted boxes. This classifier is applied to the validation set to identify and remove "not cell" boxes. CDM calculates the F1 score on the remaining validation boxes. The $\alpha$ value that yields the highest validation F1-score is selected as optimal.

Once the optimal $\alpha$ is determined, CDM applies the adjusted cutoff with this value to the testing set and uses the corresponding trained cell classifier to classify the detected bounding boxes. After removing all the bounding boxes classified as "not cell", CDM returns the remaining bounding boxes as the final detected cells.

\cref{fig:CDM_process} provides a visual comparison of the detection results obtained by different methods. Each row represents a zoomed-in part of an original AS-OCT image, and each red square represents the detected bounding box (all are in the same fixed size of $h \times w$). The first column shows the original image, the second displays the bounding boxes labeled by experts, the third presents the bounding boxes identified by the thresholding algorithm, the fourth showcases the bounding boxes after applying the adjusted cutoff, and the fifth depicts the final CDM results after discarding "not cell" objects.

The traditional thresholding algorithm (third block) fails to detect some cells compared to the expert labels (second block). The adjusted cutoff (fourth block) includes more candidate bounding boxes, but some could be false positives, such as the bottom-right box. The cell classifier determines whether each bounding box contains a "real cell" and discards "not cell" objects, resulting in the final CDM results (fifth block).

CDM enables the detection of a higher number of candidate cells while maintaining accuracy by utilizing a cell classifier to filter out false positives.

\subsection{Implementation Details}
\subsubsection{CSM}
We empirically set $R_{AS}$ to 0.7, set $N_p$ to 2 and define the offsets as $(\lambda^X_1, \lambda^Y_1) = (0, 0.05 \times \text{image width})$ and $(\lambda^X_2, \lambda^Y_2) = (0, -0.05 \times \text{image width})$, positioning the point prompts slightly to the left and right of the anterior segment's centroid.
\subsubsection{CDM}
In this study, $h$ and $w$ are both set to 20, while $H$ and $W$ are 1598 and 1465, respectively. For the adjusted cutoff algorithm, we select the Isodata thresholding method as the base algorithm, as it is commonly used for cell detection in AS-OCT images \cite{lu2020quantitative, etherton2023quantitative, agarwal2009high, baghdasaryan2019analysis}. $\beta_{max}$ is set to 25, as objects with an area larger than this value are unlikely to be cells \cite{lu2020quantitative, etherton2023quantitative}. $\beta_{min}$ is set to 1 to include as many candidate cells as possible. The values for $\tau_{max}$, $\tau_{min}$, and $\tau_{step}$ are set to 0.7, 0.99, and 0.01, respectively.

The cell classifier is a neural network with a hidden layer architecture of [512, 128, 64] designed to classify each bounding box as a real cell or not. The classifier is trained using the cropped images from the training and validation sets, with binary cross-entropy loss as the objective function. To prevent overfitting, the training is halted if the validation loss does not decrease for 30 consecutive epochs.

\section{Experiment}
\label{sec:experiment}
\subsection{Dataset and Ethics Statement}
The AS-OCT datasets were obtained from the Imaging in Childhood Uveitis studies \cite{akbarali2021imaging, etherton2023quantitative, solebo2024establishing}, which include patients from specialist children's and eye hospitals in London, UK. To annotate the dataset, we utilized the ongoing Citizen Science project \cite{jones2018crowd} hosted on the Zooniverse\footnote{https://www.zooniverse.org/} platform. Six ophthalmic clinicians used the platform\cite{solebo2024eyes} to label the AC area and cell location in the dataset. Afterward, a senior ophthalmologist verified the annotations. In total, we obtained 1,376 annotated images, with 630 images for AC segmentation and 746 images for cell detection.

This study received the necessary research approvals from UK’s Health Research Authority (South Central Oxford B Research Ethics Committee, reference 19/SC/0283:SA02). Formal written consent was obtained from participants.

\subsection{Evaluation Metrics}
For AC segmentation, we employ the Intersection over Union (IoU) and the Dice coefficient. IoU measures the overlap between the predicted and ground truth masks, while the Dice coefficient quantifies their similarity:
$$IoU = \frac{TP_{pixel}}{TP_{pixel} + FP_{pixel} + FN_{pixel}}$$
$$Dice = \frac{2 * TP_{pixel}}{2 * TP_{pixel} + FP_{pixel} + FN_{pixel}}$$
where $TP_{pixel}$, $FP_{pixel}$, and $FN_{pixel}$ are the number of true positives (correctly segmented pixels), false positives (incorrectly segmented pixels), and false negatives (ground truth pixels missed by the segmentation), respectively.

For cell detection, we use precision, recall, and F1-score at both the image and bounding box levels. A predicted bounding box is considered correct if it contains a ground truth cell location. Once a ground truth cell is associated with a predicted bounding box, it is marked as identified and removed from consideration for future predicted bounding boxes. For the $i^{th}$ image, we define $TP_{box}^i$, $FP_{box}^i$, and $FN_{box}^i$ as the number of true positives (correct predicted boxes), false positives (predicted boxes without ground truth points), and false negatives (unidentified ground truth cells), respectively. Precision and recall for the $i^{th}$ image are:
$$Precision_{img}^i = \frac{TP_{box}^i}{TP_{box}^i+FP_{box}^i}$$
$$Recall_{img}^i = \frac{TP_{box}^i}{TP_{box}^i+FN_{box}^i}$$
For the images where ground truth cell points are present but the method cannot identify any, or for the images without any ground truth cells but with predicted cells, both precision and recall are set to 0. For image-level evaluation, we calculate the average precision ($Precision_{img}$) and recall ($Recall_{img}$) across all images. They are then used to calculate the F1-score:
$$F1_{img}=\frac{2*Precision_{img} * Recall_{img}}{Precision_{img}+Recall_{img}}$$

To evaluate the correctness of all predicted bounding boxes (individual cell detection), we report the bounding box level precision, recall, and F1-score:
$$Precision_{box}=\frac{\sum_i TP_{box}^i}{\sum_i TP_{box}^i+\sum_i FP_{box}^i}$$
$$Recall_{box}=\frac{\sum_i TP_{box}^i}{\sum_i TP_{box}^i +\sum_i FN_{box}^i}$$
$$F1_{box}=\frac{2*Precision_{box} *Recall_{box}}{Precision_{box}+Recall_{box}}$$

\subsection{Experiment and Results}

\begin{table}
\caption{Comparison of segmentation performance between CSM and baseline models.}
\label{tab:res_chamber_seg}
\centering
\begin{tabular}{ccc}
\hline
\textbf{$Methods$} & \textbf{$IoU$}           & \textbf{$Dice$} \\
\hline
DeepLabV3+ \cite{chen2018encoder} & 93.51\% & 96.53\% \\
PSP Net \cite{zhao2017pyramid} & 93.76\% & 96.71\% \\
Swin-Unet \cite{cao2022swin} & 95.65\% & 97.91\% \\
UNet++ \cite{zhou2018unet++} & 95.87\% & 97.92\% \\
UNet \cite{ronneberger2015u} & 95.98\% & 97.86\% \\
nnUNet \cite{isensee2021nnu} & 96.05\% & 98.03\% \\
CSM (\textbf{Ours}) & \textbf{96.41\%} & \textbf{98.21\%} \\
\hline
\end{tabular}
\end{table}

The data were randomly assigned to training, validation, and testing sets with ratios of 40\%, 10\%, and 50\%, respectively. For the chamber segmentation task, the sizes of the training, validation, and testing sets are 252, 63, and 315, respectively. For the cell detection task, the sizes are 298, 75, and 373, respectively. The experiment was repeated 10 times, with different training, validation, and testing data in each iteration.

We evaluate the performance of CSM on the segmentation dataset and compare it with current SOTA segmentation models, including UNet \cite{ronneberger2015u}, UNet++ \cite{zhou2018unet++}, nnUNet \cite{isensee2021nnu}, Swin-Unet \cite{cao2022swin}, PSPNet \cite{zhao2017pyramid}, and DeepLabV3+ \cite{chen2018encoder}. The IoU and Dice coefficient results are presented in \cref{tab:res_chamber_seg}. The results demonstrate that CSM outperforms all baseline models, achieving a high IoU of 96.41\% and a Dice coefficient of 98.21\%.

\begin{table}
\caption{Comparison of cell detection methods at the image level.}
\label{tab:res_cell_detection_img}
\centering
\scriptsize
\begin{tabular}{cccc}
\hline
\textbf{$Methods$} & \textbf{$Precision_{img}$} & \textbf{$Recall_{img}$} & \textbf{$F1_{img}$} \\ 
\hline
Faster-RCNN \cite{ren2016faster} & / & / & / \\ 
Mask-RCNN \cite{he2017mask} & / & / & / \\ 
Cascade R-CNN \cite{cai2018cascade} & / & / & / \\
EfficientDet-D7 \cite{tan2020efficientdet} & / & / & / \\
DETR \cite{carion2020end}  & / & / & / \\ 
Thres(5) & 75.95\% & 48.74\% & 59.37\% \\ 
Thres(4) & 81.37\% & 56.87\% & 66.94\% \\ 
Thres(3) & 86.19\% & 66.39\% & 75.00\% \\ 
Thres(2) & \textbf{89.26\%} & 75.60\% & 81.86\% \\ 
Thres(1) & 81.66\% & 83.78\% & 82.70\% \\ 
CDM (w/o filter) & 42.15\% & \textbf{94.93\%} & 57.13\% \\ 
CDM (\textbf{Ours}) & 87.82\% & 90.29\% & \textbf{89.02\%} \\ 
\hline
\end{tabular}
\end{table}
We evaluate the performance of CDM on the cell detection dataset and compare it with the commonly used thresholding algorithm (i.e., Isodata) and with current SOTA object detection models, including Faster-RCNN \cite{ren2016faster}, Mask-RCNN \cite{he2017mask}, Cascade R-CNN \cite{cai2018cascade}, EfficientDet-D7 \cite{tan2020efficientdet}, and DETR \cite{carion2020end}. The evaluation metrics are reported at both the image level (\cref{tab:res_cell_detection_img}) and the bounding box level (\cref{tab:res_cell_detection_box}).

In \cref{tab:res_cell_detection_img} and \cref{tab:res_cell_detection_box}, "Thres($\beta_{min}^{*}$)" indicates that objects with an area less than $\beta_{min}^{*}$ pixels are not considered as cells after the image is processed by Isodata. Same as CDM, objects detected by Isodata with an area larger than $\beta_{max}$ (=25) are also removed. "CDM" represents our proposed model, which uses adjusted cutoff with optimal $\alpha$ to process the images in the testing set, followed by the cell classifier to remove "not cell" objects. "CDM (w/o filter)" shows the results of only the adjusted cutoff (with optimal $\alpha$) without the filtering "not cell" boxes. The "/" symbol indicates that the model failed to provide any bounding boxes on the validation set during training.

The results in \cref{tab:res_cell_detection_img} and \cref{tab:res_cell_detection_box} demonstrate that CDM achieves the best F1-score overall, outperforming the best baseline method by 6.32\% and 6.55\% at the image and bounding box levels, respectively. This highlights the effectiveness of our proposed approach in detecting cells accurately.

\begin{figure}[t]
\centerline{\includegraphics[width=\columnwidth]{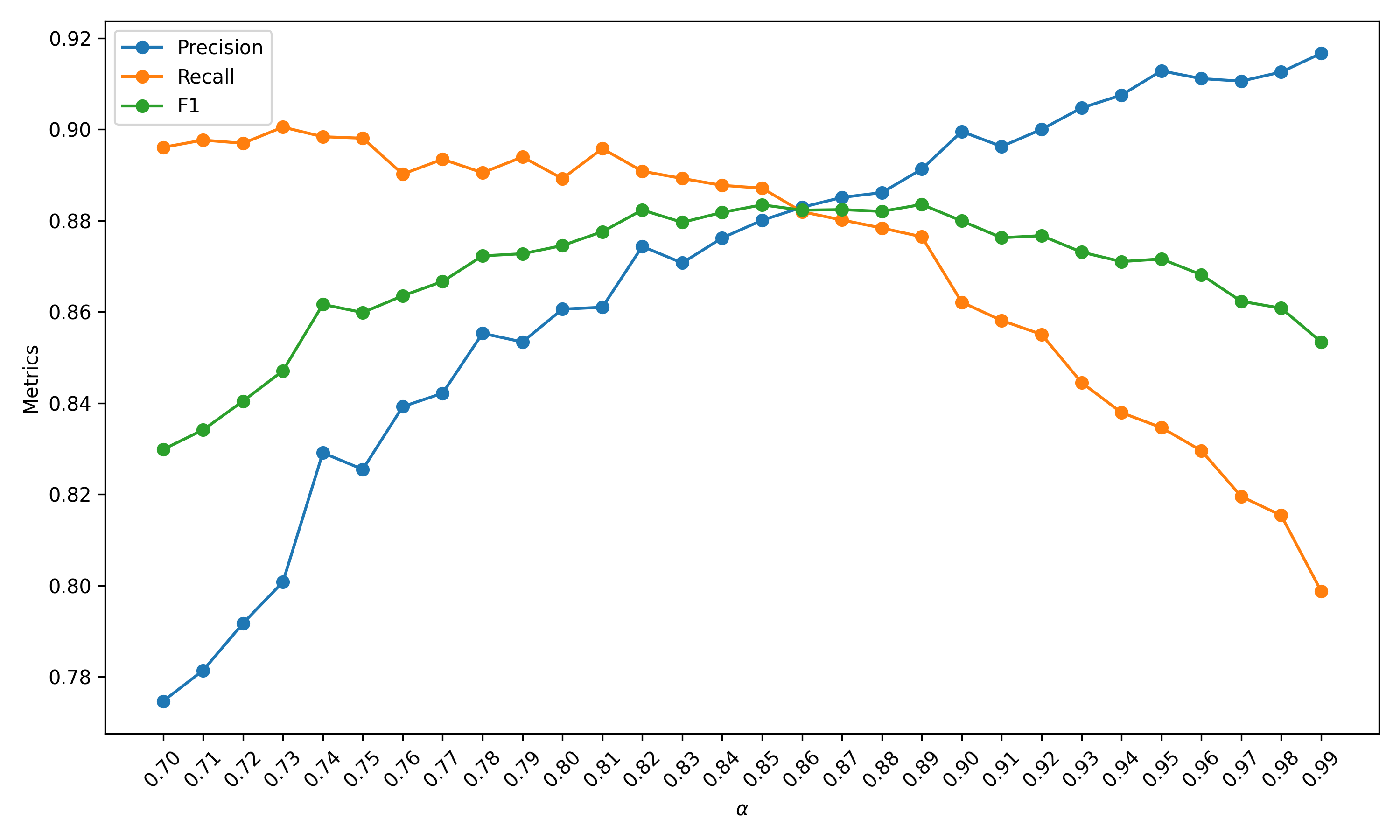}}
\caption{Trends of evaluation metrics with the change of $\alpha$ in validation set.}
\label{fig:metrics_plot_val}
\end{figure}

To illustrate how different $\alpha$ values (from 0.7 to 0.99 with a step size of 0.01) affect the model's performance, we present the change in $Precision_{box}$, $Recall_{box}$, and $F1_{box}$ on the validation set as $\alpha$ varies, as shown in \cref{fig:metrics_plot_val}. The figure shows that precision increases as the $\alpha$ approaches 0.99, while recall increases as the $\alpha$ value moves away from 0.99. Interestingly, the F1-score first increases to a peak and then decreases as the $\alpha$ moves away from 0.99. This peak value represents the optimal $\alpha$ used in the testing set, which balances precision and recall to achieve the best overall performance.
\begin{table}
\caption{Comparison of cell detection methods at the bounding box level.}
\label{tab:res_cell_detection_box}
\centering
\scriptsize
\begin{tabular}{cccc}
\hline
\textbf{$Methods$} & \textbf{$Precision_{box}$} & \textbf{$Recall_{box}$} & \textbf{$F1_{box}$} \\ 
\hline
Faster-RCNN \cite{ren2016faster} & / & / & / \\
Mask-RCNN \cite{he2017mask} & / & / & / \\
Cascade R-CNN \cite{cai2018cascade} & / & / & / \\
EfficientDet-D7 \cite{tan2020efficientdet} & / & / & / \\
DETR \cite{carion2020end}  & / & / & / \\ 
Thres(5) & \textbf{98.64\%} & 44.99\% & 61.78\% \\
Thres(4) & 98.26\% & 53.03\% & 68.87\% \\
Thres(3) & 97.26\% & 61.40\% & 75.27\% \\
Thres(2) & 93.59\% & 70.58\% & 80.47\% \\
Thres(1) & 73.89\% & 79.96\% & 76.79\% \\
CDM (w/o filter) & 16.34 \% & \textbf{93.96\%} & 26.91\% \\
CDM (\textbf{Ours}) & 86.88\% & 87.23\% & \textbf{87.02\%} \\
\hline
\end{tabular}
\end{table}

\section{Discussion}
\label{sec:discussion}
\subsection{Chamber Segmentation}
CSM leverages SAM to delineate the AC area in AS-OCT images in a fully automatic manner, achieving superior performance without the need for annotated data. In contrast, the baseline SOTA segmentation models, including UNet \cite{ronneberger2015u}, UNet++ \cite{zhou2018unet++}, nnUNet \cite{isensee2021nnu}, Swin-Unet \cite{cao2022swin}, PSPNet \cite{zhao2017pyramid}, and DeepLabV3+ \cite{chen2018encoder}, rely on manual annotations for training.

The success of CSM lies in its integration of SAM with our HPG algorithm. While previous studies \cite{ma2024segment, gu2024lesam, jiang2024glanceseg, wu2023medical, zhang2023customized} have adapted SAM for medical image segmentation tasks, they still require users to manually provide prompts or fine-tune the pre-trained model. In contrast, CSM eliminates these requirements, and achieving a fully automatic and zero-shot approach, making it more practical and efficient for clinical applications. The HPG algorithm automatically generates appropriate prompts based on the characteristics of the AS-OCT image, which are then fed into the pre-trained SAM model to segment the desired region without additional training or fine-tuning.

CSM highlights the potential of prompt engineering based on heuristic rules to adapt large foundation models to specialized domains and tasks. Our study demonstrates the effectiveness of SAM in accurately segmenting anatomical structures like the AC when provided with appropriate prompts. This opens up possibilities for future research to explore optimizing prompt generation for other structures, further extending the applicability of this approach.

\subsection{Cell Detection}
CDM demonstrates superior performance in cell detection within the segmented AC area, outperforming both the commonly-used thresholding algorithm and current SOTA deep learning object detection models.

Initially, we attempted to use SOTA deep learning object detection models, including Faster-RCNN \cite{ren2016faster}, Mask-RCNN \cite{he2017mask}, Cascade R-CNN \cite{cai2018cascade}, EfficientDet-D7 \cite{tan2020efficientdet}, and DETR \cite{carion2020end}, for this task. However, our experiments (\cref{tab:res_cell_detection_img} and \cref{tab:res_cell_detection_box}) showed that these models failed to detect any cells. The challenge lies in the small size of the target objects, with each bounding box occupying only $20 \times 20$ pixels, representing less than 0.02\% of the total image area ($1598 \times 1465$ pixels). This makes the training process extremely difficult. 

One possible solution is to split the high-resolution AS-OCT images into smaller image blocks and apply the models to these sub-images. However, this approach presents two significant challenges. First, determining the optimal cropped size for the image blocks is difficult. Second, splitting images into different blocks may result in individual cells being divided across adjacent image blocks. Consequently, we concluded that deep learning object detection models are not well-suited for detecting inflammatory cells in high-resolution AS-OCT images.

We then explored alternatives by applying thresholding algorithms, which have been commonly used to identify cells as they can identify very small object in an image \cite{keino2022automated, lu2020quantitative, etherton2023quantitative, sharma2015automated, agarwal2009high, rose2015aqueous, baghdasaryan2019analysis, kang2021development}. However, our analysis (\cref{tab:res_cell_detection_img} and \cref{tab:res_cell_detection_box}) revealed that while the widely used thresholding-based technique (i.e., Isodata) exhibit high precision in cell detection, they suffer from relatively low recall, potentially overlooking a significant number of actual cells present in the images.

\cref{tab:res_cell_detection_box} shows the evaluation results based on the correctness of all individual detections (i.e., at the bounding box level). When using a strict filtering strategy in "Thres($\beta_{min}^{*}$)", where objects detected by the Isodata algorithm with an area less than 5 pixels ($\beta_{min}^{*}=5$) are not considered as cells, the highest precision of 98.64\% can be achieved. However, this comes at the cost of a very low recall of 44.99\%, indicating that mroe than half of the cells are missed. When decreasing $\beta_{min}^{*}$ to 1, thereby considering more objects as cells, the recall increases to 79.96\%, but the precision drops dramatically to 73.11\%. Across all $\beta_{min}^{*}$ settings, the F1-scores, which balance precision and recall, remain below 81\%, indicating their inability to accurately detect the location of cells in AS-OCT images.

Although previous studies reported a strong correlation between the number of detected cells and the SUN score (human count using biomicoscopy), they did not verify the correctness of the detection. Our study suggests that these correlations may have been constrained by incomplete cell detection. This indicates that while the cells detected in these studies might have been sufficient to establish a correlation with clinical scores or manual counts, the detection likely overlooked a considerable proportion of cells, leading to an underestimation of the true cell population, and an imprecision which could limited future clinical utility of this modality.

To address these limitations, we developed CDM, which introduces two key components: the adjusted cutoff and a cell classifier. The adjusted cutoff lowers the threshold computed by the thresholding algorithm, forcing it to include more candidate cells. As shown in \cref{tab:res_cell_detection_box}, "CDM (w/o filter)" (i.e., adjusted cutoff) achieves a recall of 93.96\%, a significant improvement over its base thresholding algorithm. However, this comes at the cost of a dramatically reduced precision of 16.34\% due to the introduction of many false positives. Then, CDM employs a cell classifier to filter out false positives.

Across all images (\cref{tab:res_cell_detection_img}), the precision, recall, and F1-score of CDM are 87.82\%, 90.29\%, and 89.02\%, respectively, increasing by 6.16\%, 6.51\%, and 6.32\% compared to the baseline with the highest F1-score. Across all predicted bounding boxes (\cref{tab:res_cell_detection_box}), these metrics of CDM are 86.88\%, 87.23\%, and 87.02\%, respectively. In comparison, the baseline with the highest F1-score achieves 93.59\%, 70.58\%, and 80.47\% for these metrics. The F1-score is 6.55\% higher for CDM, demonstrating its superior overall performance. Although the precision of CDM is 6.71\% lower than the baseline, the recall increases substantially by 16.65\%, indicating that CDM can identify a significantly higher number of cells.

We further explored the impact of different $\alpha$ values on the model's performance. \cref{fig:metrics_plot_val} shows that precision increases as $\alpha$ approaches 0.99, while recall increases as $\alpha$ moves downward, with the F1-score reaching a peak at an intermediate $\alpha$ value, which represents the optimal setting. As $\alpha$ decreases, more candidate cells are included, potentially causing the cell classifier to make more mistakes, resulting in a decrease in precision as recall increases. This indicates that the choice of $\alpha$ needs to balance the number of cells it can detect and the mistakes the classifier will make. CDM automatically selects the optimal $\alpha$ value from the training and validation sets, effectively increasing the recall of cell detection while maintaining reasonable precision levels, ensuring that CDM provides an improved detection of inflammatory cells within AS-OCT images.

\subsection{Limitations and Future Directions}
While ACCDor demonstrates superior performance for AC segmentation and cell detection task, there are limitations to acknowledge. One key challenge is the current lack of standardized benchmark datasets for the tasks. Facilitating the creation of such datasets remains an important goal for future research. Another limitation is that AS-OCT images may contain artifacts or have limited
quality \cite{patel2024quality}, which can compromise ACCDor’s performance. To mitigate this problem in clinical implementation, pre-processing methods can be used to exclude low-quality images \cite{chen2023automated}, as they have limited diagnostic value.

Despite these limitations, ACCDor shows potential as a fully automated framework for AC segmentation and cell detection in AS-OCT images. It opens up new avenues for exploring the spatial distribution of inflammatory cells within the AC, which could deepen our understanding of disease pathogenesis and progression, potentially informing personalized treatment strategies. 

Building upon ACCDor, future research could investigate real-time cell detection, where cells are highlighted on AS-OCT scans during image acquisition, providing immediate feedback to guide clinical decisions. In addition, incorporating temporal information across longitudinal AS-OCT scans is another promising direction. Spatio-temporal models that detect cells and model their trajectories over time may provide a clearer picture of treatment response. Furthermore, combining cell detection outputs with other clinical variables and imaging modalities could lead to more accurate predictions of disease outcomes and a more comprehensive view of ocular health.

\section{Conclusion}
\label{sec:conclusion}
Previous studies for detecting cells in AS-OCT images may have limitations in the effectiveness of detecting cells and the reliability of their detection results. To address these, we propose ACCDor, which consists of two modules: CSM for segmenting the AC area and CDM for localizing cells within the segmented region. Experimental results demonstrate that CSM outperforms SOTA segmentation models in a zero-shot manner, while CDM achieves superior cell detection performance compared to baselines. Notably, this study reveals that previous studies that used thresholding-based methods to detect cells may overlook a significant number of cells, even if they reported strong correlation between the detected cell count and clinical scores. ACCDor provides a more accurate cell detection solution, paving the way for improved disease monitoring and personalized treatment strategies for anterior uveitis patients. We believe the proposed methodology and insights could serve as a valuable perspective for future research in automated analysis of AS-OCT images and have the potential to enhance clinical decision-making in the management of anterior uveitis.

\scriptsize
\bibliographystyle{unsrt}

\end{document}